\documentstyle[preprint,eqsecnum,epsfig,prd,aps,floats]{revtex} 

\begin{document}

\newcommand{\nl}{\nonumber \\}
\newcommand{\eq}[1]{Eq.~(\ref{#1})}

\preprint{\vbox{ \tighten {
		\hbox{MIT-CTP-2945}
		\hbox{PUPT-1915}
		}  }}  
  
\title{Thick domain walls and singular spaces}
  
\author{ Martin Gremm\thanks{email: gremm@feynman.princeton.edu }\thanks{ 
On leave of absence from MIT, Cambridge, MA 02139}} 
  
\address{Joseph Henry Laboratories, Princeton University, Princeton, NJ 08544}
 
\maketitle  
 
\begin{abstract} 
\tighten{ 
We discuss thick domain walls interpolating between spaces with naked
singularities and give arguments based on the $AdS$/CFT correspondence why
such singularities may be physically meaningful. Our examples include thick
domain walls with Minkowski, de Sitter, and anti-de Sitter geometries on the
four-dimensional slice. For flat domain walls we can solve the equivalent
quantum mechanics problem exactly, which provides the spectrum of graviton
states. In one of the examples we discuss, the continuum states
have a mass gap. We compare the graviton spectra with expectations 
from the $AdS$/CFT correspondence and find qualitative agreement. We also 
discuss unitary boundary conditions and show that they project out all 
continuum states.
} 
\end{abstract} 
 
\newpage

\section{Introduction} 
 
Domain walls have recently attracted renewed attention, after it was pointed
out in \cite{RS} that four-dimensional gravity can be realized on a thin wall
connecting two slices of $AdS$ space. From the point of view of a
four-dimensional observer on the domain wall the spectrum of gravity
consists of a massless graviton and a tower of Kaluza-Klein modes with
continuous masses. It was shown in \cite{RS} that the KK modes give a subleading
correction to the gravitational interaction between two test masses on the
domain wall. In the thin wall setup there
is only gravity in the bulk, and the only five-dimensional space that can
appear is $AdS$\footnote{Of course one can also consider slices of
$dS$ or Minkowski space, but they do not yield a four-dimensional graviton.}. 
On the other hand, in supergravity or string theory one expects to have other 
bulk fields including scalars.

The original proposal of \cite{RS} has been generalized in several directions. 
One generalization involves turning on a cosmological constant on the domain
wall \cite{bent,cosm}, which results in time-dependent cosmological scenarios. 
Other extensions include higher dimensional embeddings \cite{csaki}, 
models with a mass gap for the continuum modes \cite{bs,sn}, and
realizations of domain walls in gravity coupled to scalars
\cite{mark,gubs,gremm,josh}. There is also an extensive literature on 
supergravity domain walls \cite{bc1,sugra,bc2},
but so far the construction of \cite{RS} has not been realized in supergravity.
In fact it was shown to be impossible in any of the known five-dimensional 
supergravities \cite{bc1,bc2,nogo}.

The thin wall construction of \cite{RS} has the disadvantage that the 
curvature is singular at the location of the wall. This problem can be avoided
if gravity is coupled to a scalar field. By choosing a suitable potential
for the scalar, we can readily generate smooth domain wall solutions
\cite{gubs,gremm,josh} that interpolate between two $AdS$ spaces. However, once
we have a scalar in the bulk, other space-times besides $AdS$, $dS$, and 
Minkowski space can appear. In this note we will study some examples of such
spaces.
Specifically, we consider a class of thick domain walls in gravity coupled to 
scalars, that interpolate between spaces with naked singularities
instead of regular $AdS$ horizons.  Normally
such spaces would be discarded as unphysical, but in this context there are
reasons to believe that considering these spaces may be meaningful.

One reason
for thinking so comes from the recent proposal \cite{witten} (see also
\cite{gubsads,ami}) that five-dimensional bulk
gravity in the thin domain wall case \cite{RS} has an equivalent description
in terms of a cut-off four-dimensional CFT on the domain wall, very much in the
spirit of the $AdS$/CFT correspondence \cite{maldacena}. The details of 
this correspondence are rather unclear at present. For instance, it is not
clear how to identify the CFT in the non-supersymmetric purely five-dimensional
setup of \cite{RS}, or how to match operators and KK modes. It is also unclear
how to impose a sharp cutoff on the
CFT that preserves four-dimensional Poincare invariance. However, leaving these
considerations aside, 
we can freely borrow results from the $AdS$/CFT literature on RG flows in 
five-dimensional supergravity \cite{RGflow}. In RG flows to non-conformal
theories (see e.g.~\cite{noncft,zaffaroni}) the $AdS$ horizon gets replaced with
a naked singularity. This singularity is physical in the sense that the 
singular behavior corresponds to strong coupling effects like confinement
or screening in
the boundary theory. Since the non-conformal boundary theory makes sense in the
infrared, the singular behavior of the metric must be resolved, either by
lifting to ten dimensions \cite{warner} or via string theory. Unfortunately
we are not aware of any criterion that tells us exactly which type of naked
singularity has a physical interpretation. We will simply assume that the
singularities in the space we consider can appear in RG flows to
non-conformal theories. In the last section we will discuss the validity of
this assumption. If our singularities are physical, we can think of our
five-dimensional space-times as four-dimensional 
gravity coupled to a non-conformal field theory. Such theories are well defined,
which provides a justification for considering this type of singular space.
We will give a more detailed discussion of these ideas in the last section.

A second argument for considering spaces that end in singularities comes
from analyzing the spectrum of gravity from a four-dimensional point of view. 
In the original setup \cite{RS} there was a single massless graviton and 
a continuous tower of KK states. The KK states couple to matter on the 
thin domain wall
and cause small violations of four-dimensional energy and momentum conservation.
This violation of conservation laws also occurs in the presence of a naked
singularity. The traditional point of view \cite{gell} posits that spaces
with naked singularities are physically acceptable only if one imposes
boundary conditions that guarantee four-dimensional energy and momentum 
conservation. These boundary conditions are usually referred to as unitary
boundary conditions. Note that this point of view is rather different than
the $AdS$/CFT inspired approach described above. In the latter case we want 
energy and momentum to leak out into either the $AdS$ horizon in the setup of
\cite{RS},
or into the naked singularities we discuss here. This leakage corresponds to 
four-dimensional gravitation exciting the degrees of freedom of the 
non-conformal field theory. Nonetheless, we can impose unitary boundary
conditions and analyze the spectrum of the KK modes in that case. It turns
out that these boundary conditions remove the continuum part of the KK spectrum
for the models discussed here and in \cite{RS,gremm}.
The theory of the discrete part of the spectrum is unitary, because these
modes die off rapidly enough as we approach the singularity. 

The discussion so far involved only flat domain walls with four-dimensional
Poincare invariance. Some of the solutions we study in this note can
accommodate a constant curvature on the four-dimensional slice, turning it
into four-dimensional de Sitter or anti-de Sitter space. Such bent
domain walls have appeared previously in cosmological thin wall solutions
\cite{bent} and similar thick domain walls in four dimensions 
were discussed in \cite{gregory}.
Our solution can be viewed as a non-singular analog of the cosmological thin
wall solutions. The ambient space of bent domain walls generically has horizons
or singularities. We analyze a domain wall interpolating between two singular
spaces in some detail and also give an example of a thick domain wall which
interpolates between spaces with regular horizons. The purpose of the second
example is merely to show that such solutions exist. Unfortunately it is too
complicated for an analytical treatment. 

Apart from their relevance for cosmology, bent domain walls are interesting
because they are generic solutions of five-dimensional gravity coupled to
scalars. By generic we mean that the supersymmetry inspired first order
formalism of \cite{fo1,fo2,gubs} cannot be used to generate a solution. 
In \cite{bc1,bc2,nogo} it was shown that even for flat domain walls where this
``superpotential'' formalism is applicable, there is no simple supersymmetric
extension of the $AdS$ domain wall solutions. Since the bent domain wall
solutions cannot be obtained from any known first order formalism, they are 
probably non-supersymmetric and therefore, in a sense, generic.

This paper is organized as follows. In section II we briefly review 
gravity coupled to a scalar, and discuss a class of solutions for both
flat and bent domain walls. These domain walls interpolate between spaces
with naked singularities.
Our solutions are simple enough that we can solve the quantum mechanics 
problem exactly for flat domain walls. 
In section III we analyze the spectrum
of metric fluctuations by studying the equivalent quantum mechanics problem
with and without imposing unitary boundary conditions. 
In section IV we discuss possible implications of our results, including
various speculations on the role of thick domain walls and singular spaces
in the light of the $AdS$/CFT correspondence.

\section{Gravity coupled to scalars}

The action for five-dimensional gravity coupled to a single real scalar reads
\begin{equation}
S = \int d^4x dr \sqrt{g}\left( -\frac{1}{4} R +
\frac{1}{2}(\partial\phi)^2 - V(\phi) \right),
\end{equation}
We  will consider metrics of the form
\begin{equation}\label{m1}
ds^2 = e^{2A(r)}\left( dx_0^2 - e^{2 \sqrt{\bar\Lambda} x_0}
        \sum_{i=1}^3 dx_i^2 \right) - dr^2
\end{equation}
or
\begin{equation}\label{m2}
ds^2 = e^{2A(r)}\left( e^{-2 \sqrt{\bar\Lambda} x_3}(dx_0^2 -
        dx_1^2 -dx_2^2) - dx_3^2 \right) - dr^2,
\end{equation}
where the four-dimensional slices are de Sitter and anti-de Sitter respectively.
The equations of motion following from the action and the ansatz for the 
metric are
\begin{eqnarray}\label{eom}
\phi^{\prime\prime}+4A^\prime \phi^\prime &=&
\frac{\partial V(\phi)}{\partial\phi} \nl
A^{\prime\prime} +\bar\Lambda e^{-2A}&=& -\frac{2}{3} \phi^{\prime 2} \\
A^{\prime 2} -\bar\Lambda e^{-2A} &=&
-\frac{1}{3}V(\phi)+\frac{1}{6}\phi^{\prime 2}. \nonumber
\end{eqnarray}
The prime denotes differentiation with respect to $r$, and we have assumed that
both $\phi$ and $A$ are functions of $r$ only. The equations of motion above
were obtained using the metric \eq{m1}. Reversing the sign of $\bar\Lambda$
yields the corresponding equations of motion for \eq{m2}.

If $\bar\Lambda=0$ the four-dimensional slices in Eqs.~(\ref{m1},\ref{m2})
are Minkowski, and we can use the first order formalism of \cite{fo2,gubs}
to generate solutions. The fields and the potential can be parametrized by a
single function $W(\phi)$ as
\begin{equation}\label{fofields}
\phi^\prime = \frac{1}{2}\frac{\partial W(\phi)}{\partial\phi}, \qquad 
A^\prime = -\frac{1}{3} W(\phi), \qquad
V(\phi) = \frac{1}{8} \left( \frac{\partial W(\phi)}{\partial \phi}\right)^2
-\frac{1}{3} W(\phi)^2.
\end{equation}

For $\bar\Lambda\neq0$ there is no know first order formalism, so we need
to solve the equations of motion directly. In general it is difficult to find
tractable solutions, because highly non-linear combinations of $A(r)$ and its
derivatives appear in the equations of motion. 

For domain walls with four-dimensional de Sitter slices, the equations of 
motion provide some model independent information. Assuming we have a 
reflection symmetric domain wall we can choose $A(0)=1$ and $A^\prime(0)=0$.
The second equation in \eq{eom} then implies that $A^{\prime\prime}$ is 
negative and $A(r)$ diverges to $-\infty$ faster than for $\bar\Lambda=0$.
For $\bar\Lambda=0$ such symmetric domain walls interpolate between $AdS$
spaces which have regular horizons infinitely far away from $r=0$. For
$\bar\Lambda\neq 0$ we expect a horizon or a singularity at a finite 
distance $r=r^\ast$. For domain walls with
$AdS_4$ slices there does not seem to be a similar argument.
We discuss an example with a naked singularity at $r=r^\ast$ first.

A class of solutions with naked singularities 
is given by $A(r) = n \log( d \cos( c r))$.
Unfortunately the expressions for the scalar field and the potential are
simple only if $n=1$, so we will focus on that case. 
Other choices for $n$ are equally valid,
but there is no closed from expression for $\phi$ and the potential. 
Nevertheless these cases can be analyzed numerically.
By picking suitable units for $\bar\Lambda$ we can set $d=1$.
The complete solution to the equations of motion is then given by

\begin{eqnarray}\label{bent}
A(r) &=& \log( \cos( c r)) \nl
\phi(r) &=& \frac{1}{c}\sqrt{\frac{3}{2} (c^2-\bar\Lambda)}
	\log\left( 
	\frac{1+\tan\left( \frac{c r}{2}\right)}
		{1-\tan\left( \frac{c r}{2}\right)} \right) \\
V(\phi) &=& \frac{3}{4}
	\cosh^2\left( \frac{c \phi}{\sqrt{\frac{3}{2}
		(c^2-\bar\Lambda)}}\right)
\left( 3 \bar\Lambda + c^2 -4c^2 \tanh^2\left(\frac{c \phi}{\sqrt{\frac{3}{2}
	(c^2-\bar\Lambda)}} \right) \right).\nonumber
\end{eqnarray}
This solution has only two adjustable parameters, $c,\bar\Lambda$, which 
determine the location of the singularities, the curvature of the
four-dimensional slice,  and the thickness of the wall.
The metric with this choice of $A(r)$ has a naked singularity at
$r^\ast=\pm\pi/2c$. However, the singularity is very similar to 
the one encountered
in the $AdS$ flow to $N=1$ SYM \cite{zaffaroni}. This may indicate that it
can be resolved either by lifting the five-dimensional geometry to ten
dimensions, or by string theory. The scalar diverges at the
singularity. If we think of it as a modulus from some compactification manifold,
this divergence can indicate that the compactification manifold shrinks to zero
size or becomes infinitely large,
so that the five dimensional truncation becomes invalid. There are some
examples where singularities in five dimensions 
actually correspond to non-singular ten dimensional geometries \cite{warner}.

In the limit $c^2=\bar\Lambda$ our solution simplifies dramatically. The 
scalar vanishes and the potential becomes constant. In fact, this limit
of our solution is $dS_5$ written in unusual coordinates. Note that our
solution is valid only for $c^2\ge\bar\Lambda$. The curvature of the four
dimensional slice imposes the constraint $r^\ast \le \pi/2\sqrt{\bar\Lambda}$
on the location of the horizon.

By changing the sign of $\bar\Lambda$ we obtain a solution for a domain wall
with $AdS_4$ slices. In that case there is no constraint on the location of
the horizons, or conversely the value of $\bar\Lambda$. 

Finally we should point out that the first order formalism of \cite{fo2,gubs}
does not apply here. For instance, using \eq{fofields} we can compute
$W(\phi)$ from the expression for $\phi$. The potential computed from $W$
has the same form as the potential above, but the coefficients do not agree. 
It would be very interesting to either find a 
first order formalism for $\bar\Lambda\neq 0$ or show that it does not exist.

The solution above has the virtue that we can solve the equations of motion 
analytically, but as we pointed out, it has naked singularities at a finite
distance from the center of the domain wall. This behavior is not generic. It
is easy to pick $A(r)$ such that we get regular horizons instead of
singularities. On such example with three free parameters is
\begin{equation}
e^{A(r)} = (r^2-r^{\ast 2}) \left(
	 \frac{\sqrt{\bar\Lambda}}{2 r^\ast}
	\left(1+\frac{1}{4 r^{\ast 2}} (r^2-r^{\ast 2}) \right)
 + c (r^2-r^{\ast 2})^2 \right),
\end{equation}
but the solutions for $\phi(r)$ and the potential have to be obtained
numerically. We simply mention this example to show that such solutions
exist, but we will not investigate it further in this paper. 

In the limit $\bar\Lambda=0$ four-dimensional Poincare invariance is restored
and we can write down the solution for all $n$.
\begin{eqnarray}\label{bl0}
A(r) &=& n \log( \cos( c r)) \nl
\phi(r) &=& \sqrt{\frac{3n}{2}} \log\left(
	\frac{1+\tan\left( \frac{c r}{2}\right)}
		{1-\tan\left( \frac{c r}{2}\right)} \right) \\
V(\phi) &=& \frac{3nc^2}{4}\left(
	\cosh^2\left(\sqrt{\frac{2}{3n}} \phi\right)
	-4n\sinh^2\left(\sqrt{\frac{2}{3n}} \phi\right)
	\right)\nonumber
\end{eqnarray}
Note that in this case the solution can be parametrized by $W(\phi)$, so we
could have found it using the first order formalism of \cite{fo2,gubs}.
Since the form of $A(r)$ is the same as in the previous example, this solution
also has naked singularities at $r^\ast=\pm \pi/2c$, but unlike in the previous 
example there is no limit in which they disappear. If $n=1/4$ the potential 
is constant and near the singularity we have
$g_{00}= e^{2A} \sim \sqrt{r^\ast-r}$.
This is the behavior found in \cite{RGflow} for general flows assuming that 
the potential can be neglected near the singularity.

Since this solution has four-dimensional Poincare invariance, we expect to
find a massless graviton and a tower of KK excitations. This solution is 
simple enough that we can give a complete solution of the equivalent quantum
mechanics problem for $n=1$. The $n=2$ case is also tractable at the level of
the example in \cite{gremm}. We will comment on the differences between these
two examples in Section IV.

\section{Graviton Fluctuations}

The solutions in the previous section provide backgrounds in which the 
fluctuations of the metric exhibit interesting behavior. It is difficult
to analyze the metric fluctuations in general, since they couple to fluctuations
of the scalar field. However, it was shown in \cite{gubs} that there is a
sector of the metric fluctuations that decouples from the scalar and satisfies
a simple wave equation. Strictly speaking this is true only if the
four-dimensional slice is Minkowski space. If the four-dimensional space is
curved, there can be an extra curvature term in the equations of motion for
the metric fluctuations. We will ignore these subtleties in this section and
study solutions to the scalar wave equations. If the four-dimensional space
is flat, the agruments of \cite{gubs} indicate that we are computing the 
mass spectrum of metric fluctuations. For curved four-dimensional slices we
assume that the solutions of the scalar wave equation have qualitatively 
the same features as the metric fluctuations\footnote{We thank C.~Kennedy for
pointing out an incorrect statement in the previous version of this section.}.  

A general metric fluctuation takes the form
\begin{equation}
ds^2 = e^{2A(r)}(g_{ij}+h_{ij}) dx^i dx^j - dr^2,
\end{equation}
where we have made a gauge choice, and $g_{ij}$ is the four-dimensional $dS$,
$AdS$ or Minkowski metric. The fluctuation $h_{ij}$ is taken to be small, so
the linearized Einstein equation provides the equation of motion for it. 
As shown in \cite{gubs} the transverse traceless part of the metric
fluctuation, $\bar h_{ij}$, satisfies the same equation of motion as a
five-dimensional scalar. It turns out that transforming to conformally flat
coordinates simplifies this wave equation considerably. In terms of
$z=\int dr e^{-A(r)}$ the metric takes the form
\begin{equation}\label{cfflat}
ds^2 = e^{2A(z)} \left( g_{ij} dx^i dx^j - dz^2 \right),
\end{equation}
and if the four-dimensional slices are $dS_4$,
the transverse traceless parts of the metric fluctuation satisfy
\begin{equation}
\left( \partial_z^2 + 3 A^\prime(z) \partial_z -
\partial_{x_0}^2 - 3 \sqrt{\bar\Lambda} \partial_{x_0} +
e^{-2 \sqrt{\bar\Lambda} x_0} \sum_{a=1}^3\partial_{x_a}\right) \bar h_{ij} = 0.
\end{equation}
This equation can be simplified further by rewriting the metric fluctuation as
$\bar h_{ij} = e^{-3(A+\sqrt{\bar\Lambda}x_0)/2} \rho_k(x) \psi_{ij}(z)$, where
$\rho_k(x)$ satisfies $g^{ij}\partial_i\partial_j \rho_k(x) = - k^2 \rho_k(x)$.
Dropping the indices on $\psi$ we finally find 
\begin{equation}\label{sch}
\left( -\partial_z^2 + V_{QM} - k^2\right) \psi = 0
\end{equation}
with
\begin{equation}\label{qmpot}
V_{QM} = -\frac{9 \bar\Lambda}{4} +
\frac{9}{4} A^{\prime 2}(z) + \frac{3}{2} A^{\prime \prime}(z).
\end{equation}
Note that for $\bar\Lambda \neq 0$ there is an extra constant piece in the 
potential. This implies that the quantum mechanics problem does not 
factorize as in \cite{gubs}. The argument given there now constrains
$k^2+9\bar\Lambda/4$ to be positive. We will come back to this point when we
analyze the solutions of this Schr\"odinger equation.

These equations were written assuming that the four-dimensional slice is $dS_4$.
The analogous equation for $AdS_4$ slices can be obtained
by the analytic continuation $x_0 \to i x_3$, $x_3 \to i x_0$, and 
$\sqrt{\bar\Lambda} \to i \sqrt{\bar\Lambda}$. We are now ready to turn to
specific examples.

The metric for the solution  in \eq{bent} can be transformed to the 
conformally flat form, \eq{cfflat}, with $A(z) = - \log( \cosh(c z))$.
Using \eq{qmpot}, we find for the potential
\begin{equation}
V_{QM} = \frac{9(c^2-\bar\Lambda)}{4}-\frac{15 c^2}{4}\frac{1}{\cosh^2(c z)}.
\end{equation}
The shape of this potential is rather different than the one found in
\cite{gremm} for a thick domain wall interpolating between $AdS$ spaces. 
The most important differences are that our potential asymptotes to a 
non-zero constant for $z\to \pm\infty$ and that there is no potential barrier
separating the asymptotic region from the interior of the domain wall.
Since the potential asymptotes to a constant we will find plane wave solutions
for sufficiently large $k^2$, but these solutions are 
separated from the discrete modes by a mass gap. These general observations
can be made precise.
The Schr\"odinger equation with this potential has a general solution
\begin{eqnarray}\label{sol}
\psi &=&
a\,\, {_2F_1}\left( -\epsilon-\frac{3}{2}, 1-\epsilon+\frac{3}{2},
	1-\epsilon,\frac{1}{2}(1-x)\right) (1-x^2)^{-\epsilon/2}\nl
&&+ b\,\, {_2F_1}\left( \epsilon-\frac{3}{2}, 1+\epsilon+\frac{3}{2},
	1+\epsilon,\frac{1}{2}(1-x)\right) (1-x^2)^{\epsilon/2},
\end{eqnarray}
where $x=\tanh(c z)$ and $\epsilon^2 = - k^2/c^2 + 9(c^2-\bar\Lambda)/4c^2$.
To find the discrete part of the spectrum
we set $a=0$ to ensure that $\psi$ is regular at $z=\infty$ ($x=1$). If
$\epsilon-3/2 = -n \in {\bf Z}_0^+$, the solution is also finite as
$z\to -\infty$ ($x=-1$), so the discrete part of the spectrum is given by
$\epsilon_n = 3/2-n$, $n=0,1$, or
\begin{equation}
k_n^2 = \frac{9(c^2-\bar\Lambda)}{4} - c^2\left( \frac{3}{2}-n\right)^2,
\end{equation}
and the corresponding wave functions are
\begin{equation}
\psi_0(z) \sim \frac{1}{\cosh^{3/2}(c z)}, \quad
\psi_1(z) \sim \frac{\sinh(c z)}{\cosh^{3/2}(c z)}.
\end{equation}
For $\bar\Lambda=0$ we find the expected massless graviton with $k^2=0$ and
an excited state with $k^2=2c^2$. For $\bar\Lambda > 0$ the four-dimensional
slice is $dS$ and at least the lowest $k^2$ is negative. If the
four-dimensional metric is $AdS$ ($\bar\Lambda < 0$), all $k^2$ are positive. 
These results may appear surprising at first sight. However, we should keep 
in mind that the notion of mass is somewhat murky in $AdS$ and $dS$ spaces. 
In both of these cases, $k^2$ is a constant that appears in the separation
of variables. It should not be confused with a four-dimensional mass. If we
put $\psi_0$ and $\rho_k$ for the lowest $k^2$ into the expression for 
$\bar h_{ij}$ we find that the metric fluctuation always satisfies the
four-dimensional wave equation $D_l D^l h_{ij} = 0$, $l=0,1,2,3$. There are
several definitions of mass in $dS$ and $AdS$, so it is not clear if these
fields are massless, but fields that satisfy this wave equation never signal
an instability. It is worth mentioning that the negative values of $k^2$
are possible because the factorization argument of \cite{gubs} constrains
the combination $k^2+9\bar\Lambda/4$ to be positive.

The solutions of the Schr\"odinger equation, \eq{sch}, also include a continuous
spectrum of eigenfunctions with $\epsilon^2 \le 0$ that asymptote to plane waves
as $z\to\pm\infty$. Formally these solutions can be obtained from \eq{sol} 
by substituting $\epsilon\to i \kappa$.
For $x \to 1$ ($z\to\infty$) we find the
asymptotic behavior $\psi(z) \sim a e^{-i c\kappa z} + b e^{i c\kappa z}$,
plane waves as expected. It is easy to show that \eq{sol} also
asymptotes to a plane wave for $x\to -1$.
To summarize, the solutions to the Schr\"odinger equation, \eq{sch}, consist
of two normalizable states with discrete eigenvalues, and a continuum of
states that asymptote to plane waves at infinity.

To proceed in our discussion,
we will now specialize to the case $\bar\Lambda = 0$, so we can interpret $k^2$
as a four-dimensional mass. In this limit our solution describes a thick 
domain wall
interpolating between two spaces with naked singularities. As mentioned in
the introduction, we can appeal to the $AdS$/CFT correspondence and think
of modes propagating in the fifth direction as excitations of some 
non-conformal field theory, which should render the four-dimensional theory
unitary. We will comment on this relation in the last section. 

This is in harmony with the approach  of \cite{RS}, which is to accept small 
violations of unitarity in the theory on the four-dimensional slice at $z=0$.
In our case this theory contains a massless graviton,
one massive state, and a continuum of modes with a mass gap of size $m_{gap} = 3c/2$. At very low 
energies, none of these massive modes can be excited, and an observer at $z=0$
sees pure four-dimensional gravity. At higher energies the massive state can
be excited, giving some corrections to Newton's law, and finally at energies
larger than the gap the whole continuum of modes can be excited. Since there
is a mass gap in the theory, the corrections to Newton's law will always be
negligible at sufficiently long distances. Violations of unitarity occur only
if modes that can travel out to the singularities can be excited, i.e.~only
at energies above the mass of the lightest continuum mode.
Both the corrections to Newton's law and the way unitarity is violated
is rather different in the thin wall scenario of \cite{RS}. There the
contribution of the KK modes is
suppressed because in the quantum mechanics description they need to tunnel
through a potential barrier. The resulting suppression of these modes at $z=0$
turns out to be sufficient to make the violations of unitarity too small to
detect in present day experiments.

Another way of dealing with the violations of various conservation laws in the
four-dimensional theory is to impose unitary boundary conditions \cite{gell}.
These boundary conditions ensure that no supposedly conserved quantities 
disappear into the singularity. This approach was used in \cite{ck} to render
a specific naked singularity harmless.

We will simplify our discussion by
introducing a new massless scalar field, $\Phi$, that satisfies the same
equation of motion as the metric fluctuation. 
This scalar field should not be confused with the scalar in the solutions in
the previous section. We will discuss the unitary boundary conditions 
in terms of this scalar because that simplifies the argument somewhat. Since
the metric fluctuations and the scalar satisfy the same equation of motion, 
the results should carry over to metric fluctuations. 

For $\bar\Lambda=0$ the metric \eq{m1} has a number of Killing vectors. It will
be sufficient for our purposes to consider only the ones generating
four-dimensional translations. These Killing vectors are given
by $\xi_i^\mu = \delta_i^\mu$, where $i$ is a four-dimensional
index. To construct currents from these Killing vectors we need the stress
tensor for a massless scalar
\begin{equation}
T_{\mu\nu} = \frac{1}{2}\partial_\mu\Phi\partial_\nu\Phi-\frac{1}{2}g_{\mu\nu}
\left( \frac{1}{2} \partial_\alpha\Phi\partial^\alpha\Phi\right).
\end{equation}
The currents $J^\mu = T^{\mu\nu} \xi^i_\nu$ satisfy conservation laws of the
form
\begin{equation}\label{cons}
\frac{1}{\sqrt{g}} \partial_\mu \left( \sqrt{g} J^\mu\right) = 0,
\end{equation}
which express the conservation of four-dimensional energy and momentum.
To ensure that these quantities are conserved in the presence
of a singularity, we demand that the flux into the singularity vanishes
\begin{equation}
\lim_{z\to\infty} \sqrt{g} J^z =
\lim_{z\to\infty} \sqrt{g} g^{zz} \frac{1}{2} \partial_i\Phi\partial_z\Phi = 0.
\end{equation}
The solution for $\Phi$ take the same form as the solutions for $\bar h_{ij}$,
i.e., $\Phi \sim e^{-3A(z)/2} \psi(z)$ with $\psi$ given in \eq{sol}. 
Using the asymptotic form 
$\psi(z) \sim a \sin(c \kappa z) + b \cos(c \kappa z)$, we find for the flux
\begin{equation}
\lim_{z\to\infty} e^{3A(z)/2} \psi(z)
\partial_z \left( e^{3A(z)/2}\psi(z)\right)\sim \psi(z) \left( (3b+2a\kappa) \cos(c \kappa z) + (3 a - 2 b\kappa) \sin(c \kappa z)\right),
\end{equation}
which does not vanish for any choice of $a,b,\kappa$ except $a=b =0$.
This implies that that the unitary boundary conditions eliminate all continuum
modes from the spectrum. It is easy to check that the two discrete modes
do not generate any flux into the singularity, so the unitary spectrum consists
of these two modes. We expect similar results if we impose unitary boundary
conditions in either the thin wall setup of \cite{RS} or the thick wall
versions in \cite{gubs,gremm,josh}. In those cases only the four-dimensional
massless graviton survives, and the continuum of KK states are projected out.

This situation should be contrasted with the example encountered
in \cite{ck}. In that case the potential in the quantum mechanics problem 
diverges near the singularity. This results in an infinite tower of 
eigenfunctions with discrete eigenvalues. The potential in our example 
asymptotes to a constant near the singularity, so we get a continuum of
plane wave states. It turns out to be impossible to satisfy the no flux
condition with these solutions, which implies that all of these states are
projected out. 

We close this section with a brief comment on the solutions with $\bar\Lambda=0$
and $n>1$.
For $n=2$ the transformation to conformally flat coordinates is given by
$z = \tan(c r)/c$ and the conformal factor reads $A(z) = - \log( 1+c^2 z^2)$.
The potential in the quantum mechanics problem is given by
\begin{equation}
V_{QM} = -3 c^2 \frac{ 1-4 c^2 z^2}{ (1+c^2 z^2)^2}.
\end{equation}
This potential appeared previously in \cite{josh} and a similar potential
was discussed in detail in \cite{gremm}.  Unlike the $n=1$ potential,
this potential asymptotes to zero for large $z$. There is
one discrete bound state at threshold and a continuum of states that
asymptote to plane waves as $z\to \pm \infty$. We can repeat the analysis 
above for this potential with essentially the same result. Imposing unitary
boundary conditions eliminates the continuous spectrum, leaving only the 
four-dimensional graviton, while invoking the $AdS$/CFT correspondence allows
us to retain the continuum. In that case the entire fifth dimension gets 
reinterpreted as a non-conformal field theory on the four-dimensional slice,
and any bulk excitations should be viewed as excitations of this field theory.

\section{Discussion and speculations}

In this note we worked out an example of a thick domain wall that interpolates
between two spaces with naked singularities. Our example is simple enough
that we can compute the spectrum of the graviton KK modes exactly. It is 
possible to extend this solution to domain walls with cosmological constants
in the four-dimensional slices. These domain walls can be viewed as non-singular
analogs of the bent thin domain walls that appeared in the literature as 
cosmological extensions of the setup in \cite{RS}.

There are other reasons
for considering this type of thick wall. Thick walls with 
four-dimensional Minkowski slices can be obtained from a first order 
``superpotential'' formalism, but to find bent solutions one needs to solve
the non-linear equations of motion directly. In this sense the bent walls
are more generic than the flat examples. 

The traditional approach to rendering naked singularities harmless consists of
imposing unitary boundary conditions on modes propagating in the bulk.
If we take this approach for our solution, or for solutions of the type
studied in \cite{RS,gremm}, we project out all continuum modes, leaving
only the four-dimensional graviton and other discrete modes if any exist. 

The $AdS$/CFT correspondence offers another point of view. 
Most of this section will be devoted to comments and speculations about 
this correspondence in domain wall settings of the type studied here and
in \cite{RS,gremm}. We would like to emphasize that unlike in the original 
$AdS$/CFT correspondence \cite{maldacena,wk}, there is at present no precise
recipe for
relating five-dimensional gravity to a boundary field theory in domain wall
space-times of the type studied in \cite{RS}. Without such a recipe, our 
comments are necessarily of a very speculative nature, but we hope that 
some of them may lead to a firmer understanding of this correspondence in time.

Before discussing the thick domain walls studied here and in \cite{gremm}, let
us briefly review how the $AdS$/CFT correspondence is expected to work in the 
scenario of \cite{RS}. The setup in \cite{RS} consists of a thin domain
wall separating the horizon parts of two $AdS$ spaces. Usually the ${\bf Z}_2$
symmetry of this space-time is gauged, so that the two slices of $AdS$ are 
identified. The location of the domain wall cuts off the $AdS$ space at some
finite radial distance. Gravity in the slice of $AdS$ is expected to have a
dual description as a strongly coupled cutoff CFT on the domain wall. 

We can adopt these arguments and apply them to thick domain walls. Let us first 
consider the domain wall discussed in \cite{gremm}. It can be viewed as a 
non-singular version of the setup in \cite{RS}, since it interpolates between
the horizon parts of two $AdS$ spaces. Unlike in the thin wall case, one usually
does not mod out by the ${\bf Z}_2$ symmetry of the geometry, so we have two
independent physical $AdS$ spaces. Since the thick domain wall smoothly
connects the two slices of $AdS$, there
is no sharp cutoff as in the thin wall case. A possible interpretation of 
this is that a smooth domain wall corresponds to a soft (or softer)
cutoff in the field
theory. From the field theory perspective this is more desirable than the
sharp cutoff imposed by a thin wall. While there is no known regularization
scheme with a sharp cutoff that preserves four-dimensional Poincare invariance,
we have a candidate for a soft cutoff. Dimensional Regularization preserves 
the desired invariances and corresponds to a soft cutoff in momentum space. 
Thick domain walls may be more appealing from the field theory point of view, 
but in gravity they pose additional challenges. For instance, it is not 
clear where the four-dimensional field theory is supposed to live, since the
space does not have a boundary. This problem could potentially be cured by
orbifolding the thick domain wall, which introduces a boundary at $z=0$.
The geometry already has a ${\bf Z}_2$ 
symmetry, so orbifolding it simply identifies the two $AdS$ spaces, but the
derivative of the scalar does not vanish at $z=0$, so we need to put a source
for it on the orbifold fixed point. Orbifolding the space should not affect
our speculation that the thick domain wall corresponds to a soft cutoff in the
CFT.

We now turn to the examples studied here. The main difference is that we are
considering thick domain walls that interpolate between singular spaces. As
mentioned before, such singular spaces appear in $AdS$ flows to non-conformal
theories, so we speculate that we can replace our singular five-dimensional
geometry by a non-conformal four-dimensional field theory with a soft cutoff.
This speculation is even harder to make precise than the previous one, because
five-dimensional gravity fails near the singularity and higher dimensional 
gravity or string theory has to come to the rescue. Nonetheless, we will 
forge ahead and offer some speculations on the field theory interpretations
of the singularities we studied here. 

We will discuss the $\bar\Lambda=0$ solutions given in \eq{bl0}. For $n=1$
the spectrum consist of the massless four-dimensional graviton, an excited state
with $m= \sqrt{2}c$, and a continuum of states with masses $m\ge m_{gap}= 3c/2$.
If we assume that our singular space corresponds to a non-conformal theory
such as SYM or QCD, we can attempt to interpret this KK spectrum. A confining
theory will have a strong coupling scale, $\Lambda_{QCD}$, which sets the mass
scale for the light states in the theory. We can interpret the mass gap
found in the KK modes as the energy needed to make the lightest particle in
the non-conformal
field theory. The presence of the mass gap in our solution at least does not 
automatically rule it out. Unfortunately we do not have a good interpretation
for the single massive resonance in the spectrum. This state should not have
a field theory counterpart, since it is localized on the domain wall and does
not propagate out to the singularity. Luckily we can eliminate this state
by imposing the orbifold projection discussed above for the $AdS$ domain wall.
This provides us with a boundary and eliminates this unwanted state. If 
a version of the $AdS$/CFT correspondence can be formulated at all,
it is likely to be in the orbifolded case.

We also briefly discuss the solution for $n=2$. The equivalent quantum mechanics
problem in that case cannot be solved completely, but we can obtain enough
information to discuss this case in the light of the $AdS$/CFT correspondence. 
The spectrum in this case is very much like that found in \cite{gremm}. We
find a single massless graviton and a continuum of plane wave states with masses
starting at zero. Unlike the case studied in \cite{gremm} this space has 
naked singularities at a finite distance from the domain wall. These 
singularities imply that, if this space has a field theory interpretation, it
should be in terms of a non-conformal theory. We have already discussed the
confining case above. Since there is no mass gap in the $n=2$ case, we 
suggest that this space may correspond to a theory that is free in the 
infrared. Such a theory would have excitations with masses that are continuous
from zero.  From the original form of the $AdS$/CFT
correspondence we expect the gravity description to break down completely if
the field theory becomes weakly coupled. This makes it unlikely that the
singularity for $n=2$ can be resolved by lifting to ten dimensions. The dual
description should require string theory on some highly curved manifold. 

Unfortunately all of our speculations follow from the assumption that we
can use the $AdS$/CFT correspondence to gain some intuition about the 
domain wall space-times we studied here. 
To make any of our statements more precise we would need a formulation of the
$AdS$/CFT correspondence along the lines of \cite{wk}. This is clearly a
necessary ingredient if we want to study domain walls in singular space-times
in a more reliable and systematic way.

As this paper was nearing completion, \cite{cc1} and \cite{cc2} appeared. These
papers have no direct overlap with the results here, but they provide another
motivation for studying singular spaces. These papers discuss an intrinsically
higher dimensional approach to solving the cosmological constant problem. In
their analysis they naturally encounter spaces with singularities at finite
distances. While it is not clear if the spaces we discussed here can be 
used in that context, their results provide another piece of evidence that 
singular spaces may play an important role in domain wall universe scenarios. 

After this paper was submitted a first order formalism for bent thick domain
walls appeared in the newest version of \cite{gubs}.

\acknowledgements
It is a pleasure to thank Miguel Costa, Josh Erlich, Igor Klebanov,
Lisa Randall, Yuri Shirman, and Kostas Skenderis for comments and
helpful conversations.  
This work was supported in part by DOE grants \#DF-FC02-94ER40818 and
\#DE-FC-02-91ER40671.

{\tighten 
 
}

\end{document}